\documentclass[3p,times,procedia]{elsarticle}
\usepackage{ecrc}
\usepackage{color}

\volume{00}

\firstpage{1}

\journalname{Procedia IUTAM}

\runauth{A. Lamura and G. Gompper}

\jid{piutam}

\usepackage{amssymb}

\usepackage[figuresright]{rotating}

\begin{document}

\begin{frontmatter}



\dochead{IUTAM Symposium on Dynamics of Capsules, Vesicles and Cells in Flow}

\title{Rheological properties of sheared vesicle and cell suspensions}


\author[a]{Antonio Lamura\corref{cor1}} 
\author[b]{Gerhard Gompper}

\address[a]{Istituto Applicazioni Calcolo, CNR,
Via Amendola 122/D, 70126 Bari, Italy}
\address[b]{Theoretical Soft Matter and Biophysics, Institute of Complex 
Systems, Forschungszentrum J\"{u}lich, 52428 J\"{u}lich, Germany}

\begin{abstract}
Numerical simulations of vesicle suspensions are performed in two dimensions to study their
dynamical and rheological properties. An hybrid method is adopted, which combines a 
mesoscopic approach for the solvent with a curvature-elasticity model for the membrane. 
Shear flow is induced by two counter-sliding parallel walls, which generate a linear 
flow profile.  The flow behavior is studied for various vesicle concentrations and 
viscosity ratios between the internal and the external fluid. Both the intrinsic viscosity
and the thickness of depletion layers near the walls are found to increase with 
increasing viscosity ratio. 
\end{abstract}

\cortext[cor1]{Corresponding author. 
Tel.: +39-0805929745 ; fax: +39-0805929770.}
\end{frontmatter}

\email{a.lamura@ba.iac.cnr.it}



\section{Introduction}
A detailed understanding of the dynamical and rheological properties of suspensions of
vesicles, cells, and capsules is relevant for a wide range of applications, ranging from 
soft glasses to blood flow.\cite{rev1,rev2,rev3} 
Vesicles are deformable particles made by a closed lipidic membrane encapsulating a fluid
whose rheology strongly depends on the bending of the membrane, on the viscosity contrast
(the ratio of the viscosity $\eta_{in}$ of the embedded fluid to that $\eta_{out}$ of the
surrounding fluid), and on the concentration.
The goal of experimental, theoretical and simulation studies is to obtain a complete 
picture of vesicle dynamics under flow. Moreover, vesicles can be considered as a model 
for more complex systems such as red blood cells. Here, the main difference between
vesicles and red blood cells is that the latter have a cytoskeleton attached to the
lipid bilayer which implies a shear modulus of the composite membrane. 

In dilute systems, vesicles show tank-treading (TT), tumbling (TU), and vacillating-breathing
motion\cite{rev1} depending on shear rate and viscosity contrast $\lambda=\eta_{in}/\eta_{out}$.
In the case of a highly dilute suspension of quasi-spherical vesicles, it has been 
predicted\cite{DM1,DM2} that the intrinsic viscosity 
$\eta_I=(\eta-\eta_{out})/(\eta_{out}\phi)$, where $\eta$ is the effective suspension 
viscosity and $\phi$ the vesicle concentration, decreases in the TT phase when $\lambda$
increases, attains a minimum at the critical value $\lambda \simeq \lambda_c$ where there is the
TT-to-TU transition, and then increases when $\lambda>\lambda_c$. This result has been
checked both numerically and experimentally. While a number of 
simulations\cite{MIS10,Rah10,Zha13,Thi13,Kao14} and one experiment\cite{V08} confirmed 
this picture, another experimental study\cite{S08} and our recent numerical work\cite{noi13}
find the intrinsic viscosity to increase monotonically with increasing viscosity contrast.

In this paper, we will give an overview of the numerical results obtained on the basis of a 
two-dimensional vesicle model, which is characterized by the presence of thermal membrane 
undulations as well as of thermal noise which implies translational Brownian motion\cite{noi13}. 
Both these features are missing in the other numerical approaches. We will briefly 
discuss their role in the obtained results for $\eta_I$, which will be given
for different values of viscosity contrast, vesicle concentration, and swelling degree.
Finally, the presence of depletion layers (or vesicle-free layers) near walls, in the case of 
concentrated suspensions, will be pointed out as a function of the viscosity contrast.

\section{Model and Methods}

In this section, we outline the model and mesoscale hydrodynamics approach employed in the 
simulations.

\subsection{Solvent}

We consider a two-dimensional fluid consisting of
$N$ point-like particles of mass $m$ with positions ${\bf r}_i(t)$ and
velocities ${\bf v}_i(t)$ at time $t$ ($i=1,2,...,N$) continuous variables.
The present method is referred to as
multi-particle collision dynamics\cite{male99,kapr08,gomp09},
but it also known as stochastic rotation dynamics\cite{kiku03,yeom06}.
The evolution occurs in subsequent steps of propagation and 
collision.
The streaming of particles is performed by moving them ballistically
\begin{equation}
{\bf r}_i(t+\Delta t)={\bf r}_i(t)+{\bf v}_i(t) \Delta t \;\;\;\;\; i=1,...,N
\label{eq.prop}
\end{equation}
with $\Delta t$ the time between two collisions.
For the scattering, the system is divided into the cells of a regular
square lattice of mesh size $a$. Each of these cells is the interaction area
where an instantaneous multi-particle collision occurs.
In this step velocities are updated\cite{nogu07,goet07} as
\begin{equation}
{\bf v}_i^{new}={\bf v}_c^G + {\bf v}_i^{ran} 
- \sum_{j \in cell} {\bf v}_j^{ran} / N_c 
+ {\bf \Pi}^{-1} \sum_{j \in cell} m \left [ {\bf r}_{j,c} \times 
({\bf v}_j - {\bf v}_j^{ran}) \right ] \times {\bf r}_{i,c} \;\;\;\;\; i=1,...,N
\label{eq.coll}
\end{equation}
where ${\bf v}_c^G$ is the velocity of the center of mass of all particles 
in the cell, ${\bf v}_i^{ran}$ is a velocity chosen from a Maxwell-Boltzmann
distribution, $N_c$ is the number of particles in the cell, ${\bf \Pi}$ is the
moment-of-inertia tensor of the particles in the cell, and ${\bf r}_{i,c}$
is the position relative to the center of mass of the particles in the cell. 
The local linear and angular momenta are conserved under this 
dynamics\cite{nogu07,goet07} and the temperature is kept constant\cite{alla02}. 

The viscosity of the fluid is given by\cite{nogu08}
\begin{equation}
\eta=\frac{m}{\Delta t} 
\left [ \left ( \frac{l}{a} \right )^2 \left ( \frac{n^2}{n-1} -\frac{n}{2}\right ) 
+ \frac{1}{24} \left (n - \frac{7}{5} \right ) \right ]
\label{visc}
\end{equation}
with $n$ the average number of particles per cell, $l=\Delta t \sqrt{k_B T /m}$ 
the mean-free path, and $k_B T$ the thermal energy.
The viscosity $\eta$ is the sum of a 
kinetic contribution $\eta_{kin}$ due to particle streaming (the first term in the
square brackets), and a collisional contribution $\eta_{coll}$ 
due to particle scattering (the second term). 
Theory slightly underestimates $\eta_{kin}$ and slightly overestimates 
$\eta_{coll}$.\cite{nogu08} 
At small values of the mean free path $l/a$, as in our case, 
$\eta_{kin}$ is negligible and $\eta_{coll}$ is a little larger than the numerical 
value.
Due to this, the effective value of the fluid viscosity
is numerically evaluated by measuring the $xy$ component of the stress tensor
$\sigma_{xy}$ in a sheared system  
so that $\eta=\sigma_{xy}/\dot\gamma$ where $\dot\gamma$ 
is the shear rate.\cite{mew}

The system of size $L_x \times L_y$ is confined by two
horizontal walls which slide along the $x$ direction with velocities $\pm v_{wall}$.
Periodic boundary conditions are used along the $x$ direction. Bounce-back 
boundary conditions with virtual particles in partly filled
cells at walls are adopted to implement no-slip at walls\cite{lamura01}.
A linear flow profile $(v_x, v_y)=(\dot \gamma y, 0)$ is obtained with shear
rate $\dot \gamma= 2 v_{wall} / L_y$.

\subsection{Vesicles}

Each vesicle in two dimensions is modeled as a closed chain made of $N_v$ beads 
of mass $m_v$ connected successively.\cite{finken}
The internal potential $U$ is the sum of three contributions. 
The bending potential is 
\begin{equation}
U_{bend}=\frac{\kappa}{r_0} \sum_{i=1}^{N_v} (1-\cos \beta_{i}) ,
\label{bend}
\end{equation}
where $\kappa$ is the bending rigidity, $r_0$ is the average bond length, and $\beta_{i}$ 
is the angle between two successive bonds, controls shapes and fluctuations.
In order to keep the length of the membrane conserved, both locally and globally,
the neighboring beads are connected to each other by the harmonic potential 
\begin{equation}
U_{bond}=\kappa_h \sum_{i=1}^{N_v} 
\frac{(|{\bf r}_i-{\bf r}_{i-1}|-r_0)^2}{2 r_0^2} 
\end{equation}
where $\kappa_h$ is the spring constant and ${\bf r}_i$ is the position vector
of the $i$-th bead.  The constraint potential
\begin{equation}
U_{area} = \kappa_A \frac{(A-A_0)^2}{2 r_0^4} ,
\label{area_pot}
\end{equation}
where $\kappa_A$ is the compression modulus and $A_0$ 
is the target area of the
vesicle, is introduced to keep the internal area $A$ of the vesicle constant.

Finally, different vesicles interact via a shifted repulsive Lennard-Jones potential
\begin{equation}
U_{rep} = 
4 \epsilon \Big [ \Big(\frac{\sigma}{r}\Big)^{12}
-\Big(\frac{\sigma}{r}\Big)^{6}\Big ] 
+ \epsilon 
\label{rep_pot}
\end{equation}
truncated at its minimum $r_{cut}$.
Newton's equations of motions for the beads are integrated by using
the velocity-Verlet algorithm with time step $\Delta t_v$.\cite{allen}

\subsection{Solvent-Vesicle Coupling}

Each bead is represented by a ``rough" hard disk of radius $r_v$.
The value of $r_v$ is set in order to ensure overlap of disks and 
a full covering up of the membrane. 
Scattering occurs only when a solvent particle $i$ and a disk $j$
overlap and move towards each other so that the conditions 
$|{\bf r}_j-{\bf r}_i| < r_v$ and 
$ ({\bf r}_j-{\bf r}_i) \cdot ({\bf v}_j-{\bf v}_i) < 0$ are both satisfied. 
A second disk $k=j \pm 1$, adjacent to the $j$-th one in the same membrane,
with the smallest distance from the solvent particle $i$, is then selected.
The center of mass velocity ${\bf v}^{G}$ and the angular velocity
\begin{equation}
{\bf \omega} 
= {\bf \Pi}^{-1} \sum_{l=i,j,k} m_l {\bf r}_{l,c} \times {\bf v}_l
\end{equation}
of the $i,j,k$-particle system are computed with ${\bf \Pi}$ the
moment-of-inertia tensor and ${\bf r}_{l,c}$ the position relative to the
center of mass.  Their velocities are updated according to the rule
\begin{equation}
{\bf v}_l^{new} = 2 ({\bf v}^{G} + {\bf \omega} \times {\bf r}_{l,c})
- {\bf v}_l \;\;\;\;\; l=i,j,k
\end{equation} 
which ensures linear and angular momenta conservation.\cite{M09}

The collision step (\ref{eq.coll}) is then executed only for the fluid
particles which did not scatter with the membrane in order to prevent 
multiple collisions with the same disk in the following time steps.
Membrane disks interact with walls also by bounce-back scattering.

\subsection{Parameters}

Experimental realizations of vesicle suspensions in shear flow are characterized 
by small values of the Reynolds number $Re=\dot\gamma \rho R_0^2 / \eta_{out}$, 
with $\rho=n m/a^2$ the mass density.  It is useful
to express results in terms of dimensionless quantities such as
the reduced area $A^*=A_0/\pi R_0^2$, where
$R_0=L_0/2 \pi$ is the mean vesicle radius and $L_0$ is the membrane
length, and the reduced shear rate (or capillary number)
$\gamma^*=\dot\gamma \eta_{out} R_0^3/\kappa$.
We set $n=10$, $l_{out}=0.0064 a$ with 
$l_{in}=l_{out} \sqrt{m_{out}/m_{in}}$ (in the following the subscripts
$out/in$ will refer to quantities outside/inside the vesicle) so
that the viscosity contrast is $\lambda=\eta_{in}/\eta_{out} \simeq m_{in}/m_{out}$.
We use $L_x \times L_y=(18.95 \times 5.79) R_0$, $R_0=7.6 a$, and $v_{wall}$
such that $Re < 0.2$. Finally, 
$m_{in}$ is set to have $0.1 \leq \lambda \leq 13.0$, $m_V=3 m_{out}$, 
$N_v=480$, $\Delta t_v = \Delta t / 64$, $r_v=r_0=a/10$, $r_{cut}=a$,
$\kappa=6.58 k_B T R_0$, 
$\kappa_A= 4 \times 10^{-4} k_B T$, $\kappa_h=3 \times 10^2 k_B T$, 
$\epsilon=10 k_B T$, and $A_0$ in such a way that $A^*=0.8, 0.95$.

\section{Results}

\subsection{Suspension Viscosity}

Dilute and semi-dilute monodisperse suspensions of vesicles are first considered
with reduced shear rate  $\gamma^{*}=2.0$.  
This value of $\gamma^{*}$ is comparable to those 
used in previous studies on vesicle rheology\cite{MIS10,Zha13,Thi13,Kao14}.
The suspension viscosity $\eta$ is 
computed by measuring the $xy$ component of the stress tensor $\sigma_{xy}$ 
as $\eta=\sigma_{xy}/\dot\gamma$.\cite{mew}

\begin{figure}[t]
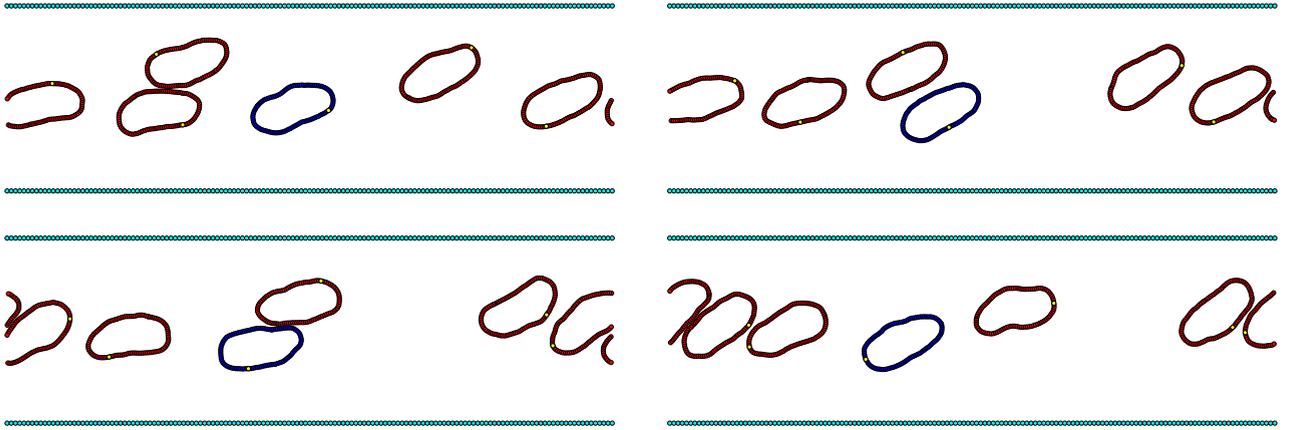
\vspace*{4pt}
\centerline{
\includegraphics[width=0.49\textwidth]{c1.epsi}
\hspace*{5mm}
\includegraphics[width=0.49\textwidth]{c2.epsi}}
\vspace*{5mm}
\centerline{
\includegraphics[width=0.49\textwidth]{c3.epsi}
\hspace*{5mm}
\includegraphics[width=0.49\textwidth]{c4.epsi}}
\caption{
Subsequent snapshots (from left to right, from top to bottom) at times $\dot{\gamma}t=224, 226, 228, 230$
of vesicle conformations in shear flow
with reduced shear rate $\gamma^*=2.0$, reduced area $A^*=0.8$,
viscosity contrast $\lambda=1.0$, and concentration $\phi=0.14$. One vesicle is in blue
for better visualization and the yellow bead indicates the tank-treading motion.
\label{fig:conf}
}\vspace*{-6pt}
\end{figure}

Figure \ref{fig:conf} shows a few typical snapshots of vesicles in shear flow. 
Vesicles displaced relative to each other in the shear-gradient direction move 
with different velocities
and can therefore collide with each other. In the TT regime, vesicles generally move
with a constant inclination angle; however, during a collision, they hug each other,
which leads to a characteristic decrease and subsequent recovery of the inclination
angle.\cite{noi13} 

\begin{figure}[t]\vspace*{4pt}
\centerline{\includegraphics[width=0.4\textwidth]{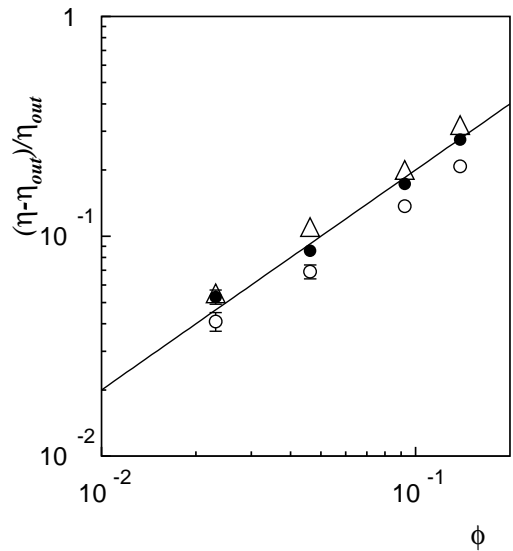}}
\caption{The quantity
$(\eta-\eta_{out})/\eta_{out}$ as a function of the concentration $\phi$
for reduced shear rate $\gamma^{*}=2.0$,
reduced area $A^{*}=0.8$, and viscosity contrasts 
$\lambda= 1.0 (\circ), 5.0 (\bullet), 9.0 (\triangle)$. 
The continuous line corresponds to the Einstein law $(\eta-\eta_{out})/\eta_{out}=2 \phi$ 
for disks in two dimensions.\cite{B81} When not visible, errors bars are comparable
with symbols size.
\label{fig:visco_rel}
}\vspace*{-6pt}
\end{figure}

\begin{figure}[t]\vspace*{4pt}
\centerline{\includegraphics[width=0.4\textwidth]{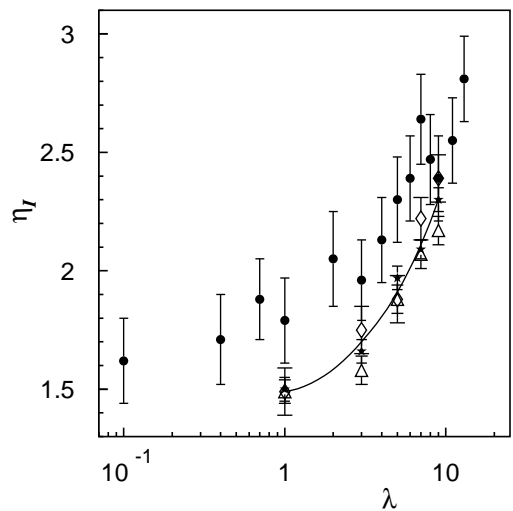}}
\caption{The intrinsic viscosity $\eta_I=(\eta-\eta_{out})/(\eta_{out} \phi$) 
as a function of the viscosity contrast $\lambda$ for reduced shear 
rate $\gamma^{*}=2.0$, reduced area $A^{*}=0.8$, and concentration
$\phi= 0.02 (\bullet), 0.05 (\Diamond), 0.09 (\triangle), 0.14 (\star)$.
The full line is the interpolations to the data for $\phi=0.14$ ($\star$). 
The tank-treading-to-tumbling 
transition occurs at $\lambda_c\simeq 3.7$ for $A^{*}=0.8$ 
in the KS theory.
\label{fig:visco_int_lambda1}
}\vspace*{-6pt}
\end{figure} 

The relative viscosity $(\eta-\eta_{out})/\eta_{out}$, displayed in 
Fig.~\ref{fig:visco_rel}, 
is a linear function of the concentration $\phi$ for reduced area $A^{*}=0.8$ and  
different values of the viscosity contrast $\lambda$, in agreement with
the linear dependence
predicted by the Einstein relation for disks in two dimensions.\cite{B81}
In Fig.~\ref{fig:visco_int_lambda1}, the intrinsic viscosity $\eta_I$ is plotted 
as a function of $\lambda$ for different concentrations with $A^{*}=0.8$.
In all the cases $\eta_I$ is observed to increase in the explored range of 
viscosity contrast.\cite{noi13}
This is at odds with some numerical results\cite{MIS10,Rah10,Zha13,Thi13,Kao14} 
and one experiment,\cite{V08} where $\eta_I$ decreases with $\lambda$, 
reaching a minimum at the TT-to-TU transition, and then increases in the 
tumbling regime, as theoretically predicted for a quasi-spherical vesicle 
in three dimensions.\cite{DM1,DM2}
In our simulations, we observe the transition from TT to TU with
increasing $\lambda$. 
For the simulated value of $A^*=0.8$, the Keller-Skalak (KS) theory\cite{KS}
indicates the TT-to-TU transition to occur at $\lambda_c \simeq 3.7$.
However, no evidence is found for the predicted dip in the intrinsic viscosity. 
We believe that this discrepancy can be traced back to the fact that our model 
includes thermal vesicle fluctuations, neglected
in the KS theory, which are known to produce a continuous crossover from
TT to TU.\cite{M09}
Our results are consistent with other 
experimental results for semi-dilute vesicle suspensions.\cite{S08} 

\begin{figure}[t]\vspace*{4pt}
\centerline{\includegraphics[width=0.4\textwidth]{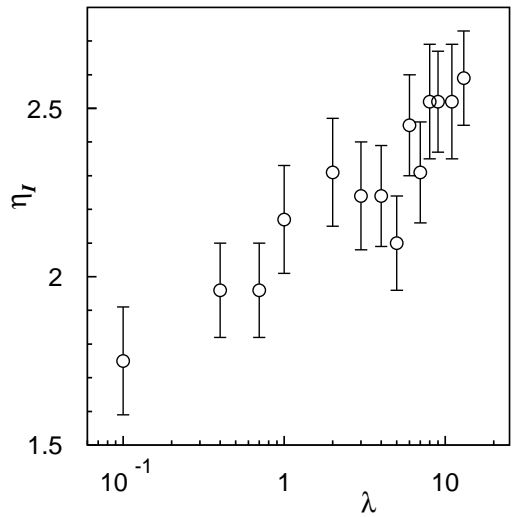}}
\caption{The intrinsic viscosity $\eta_I=(\eta-\eta_{out})/(\eta_{out} \phi$) 
as a function of the viscosity contrast $\lambda$
for reduced shear rate $\gamma^{*}=2.0$, reduced area $A^{*}=0.95$, and 
concentration $\phi= 0.03$.  The tank-treading-to-tumbling 
transition occurs at $\lambda_c\simeq 6.5$ for $A^{*}=0.95$ 
in the KS theory.
\label{fig:visco_int_lambda2}
}\vspace*{-6pt}
\end{figure}

Also the case of a single quasi-circular ($A^{*}=0.95$)
vesicle is investigated, corresponding to concentration $\phi= 0.03$. 
As for the more deflated vesicle ($A^{*}=0.8$) discussed above,
after an initial transient the vesicle moves along the channel 
while diffusing laterally up
to the longest simulated times $\sim 10^2/\dot\gamma$.
The results for $\eta_I$ are reported in Fig.~\ref{fig:visco_int_lambda2}
and confirm the general picture of intrinsic viscosity increasing
with $\lambda$.

\subsection{Depletion Layer}

\begin{figure}[t]\vspace*{4pt}
\centerline{\includegraphics[width=0.4\textwidth]{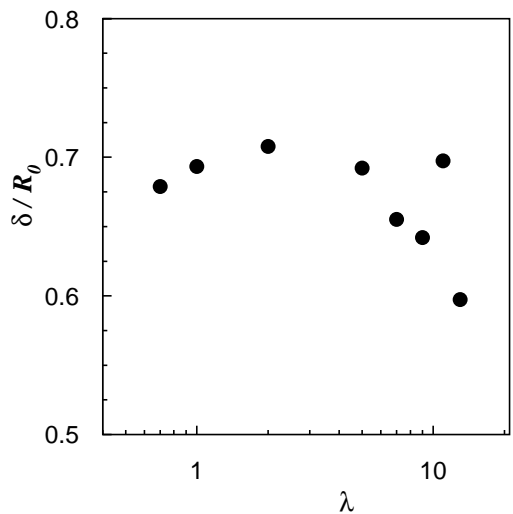}}
\caption{The average cell-free boundary layer $\delta$ 
as a function of the viscosity contrast $\lambda$
for reduced shear rate $\gamma^{*}=2.0$, reduced area $A^{*}=0.8$,  and concentration 
$\phi= 0.28$.  Error bars are comparable with symbols size.
\label{fig:vfl}
}\vspace*{-6pt}
\end{figure}

\begin{figure}[t]
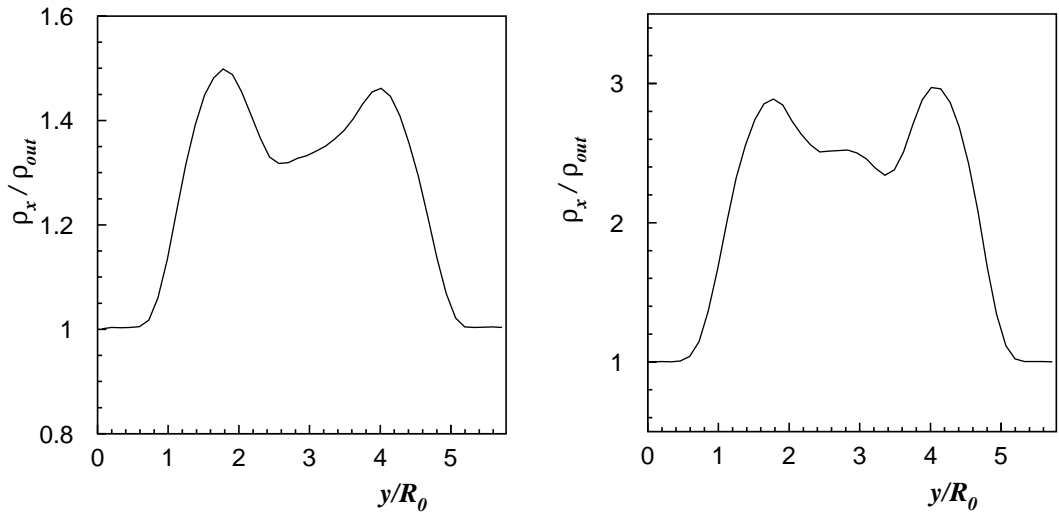
\vspace*{4pt}
\centerline{
\includegraphics[width=0.4\textwidth]{fig_cell_2.epsi}
\hspace*{5mm}
\includegraphics[width=0.4\textwidth]{fig_cell_5.epsi}}
\caption{The ratio of the 
mass density $\rho_x$, averaged along the flow direction $x$,
to the solvent mass density $\rho_{out}$ is plotted across the channel 
for reduced shear rate $\gamma^{*}=2.0$, reduced area $A^{*}=0.8$, concentration 
$\phi= 0.28$, and viscosity contrasts $\lambda= 2.0$ (left), $5.0$ (right).
\label{fig:dens_profile}
}\vspace*{-6pt}
\end{figure}

The effect of wall confinement is investigated 
for monodisperse concentrated suspensions with concentration $\phi=0.28$,
reduced area $A^*=0.8$, and reduced shear rate $\gamma^*=2.0$.
It is well known\cite{M09} that the lift force induced by the hydrodynamic 
interaction of vesicles with nearby walls produces a depletion layer near the walls,
as first observed for red blood cells in capillary flow.\cite{FL}
We have measured the average thickness $\delta$ of such vesicle-free layers
(also denoted cell-free layers in analogy with the case of red blood cells). 
Here, $\delta$ is defined as the time average of $(d_1(t)+d_2(t))/2$
where $d_1(t)$ and $d_2(t)$ are the distances at time $t$
of the two vesicle beads closest to the two walls, respectively.
The behavior of $\delta$ as a function of $\lambda$ is shown in Fig.~\ref{fig:vfl}. 
The dependence of $\delta$ on $\lambda$ is evidently non-linear. The cell-free 
layer decreases at high values of the viscosity contrast after 
reaching a maximum close to TT-to-TU transition.  The existence of cell-free 
boundary layers is also confirmed by looking at the steady mass density profiles,
averaged along the $x$ direction, which are reported 
in Fig.~\ref{fig:dens_profile} for two values of $\lambda$.

\begin{figure}[t]
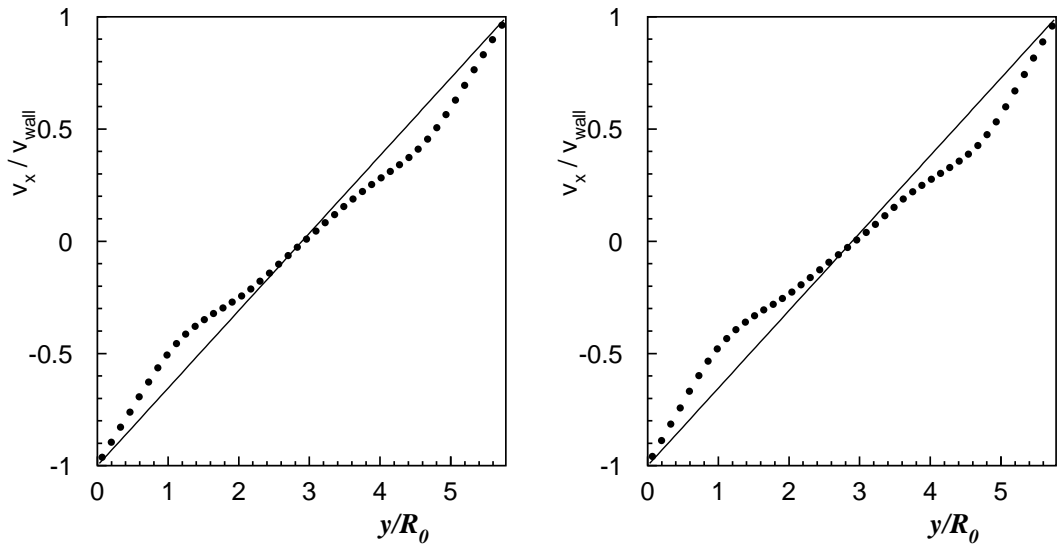
\vspace*{4pt}
\centerline{\includegraphics[width=0.4\textwidth]{fig_vel_2.epsi}
\hspace*{5mm}
\includegraphics[width=0.4\textwidth]{fig_vel_5.epsi}}
\caption{The stationary velocity profile averaged along the channel
is plotted across the channel 
for reduced shear rate $\gamma^{*}=2.0$, reduced area 
$A^{*}=0.8$, concentration 
$\phi= 0.28$,  and
viscosity contrasts $\lambda= 2.0$ (left), $5.0$ (right).
The full line shows the imposed shear rate profile.
\label{fig:velocity_profile}
}\vspace*{-6pt}
\end{figure}

This effect also appears when considering the 
corresponding steady velocity profiles, averaged along the flow direction,
which are shown in Fig.~\ref{fig:velocity_profile}.
It can be seen that the lack of vesicles, filled with a heavier fluid, 
produces lower values of mass density  close to the walls.
The effective shear rate at the center is smaller than the imposed shear rate, 
while it is larger close to the walls.
The observed behavior can be related to the interplay between the reduction of the
tilt angle in the TT regime, which favors the packing of vesicles in the channel,
and the reduction of the lift force with increasing viscosity contrast.\cite{M09}

\section{Summary and Conclusions}

We have studied the rheological properties of dilute and semi-dilute vesicle
suspensions in wall-bounded shear flow in two dimensions. We find that 
the intrinsic viscosity
is an increasing function of the viscosity contrast.
As pointed out in Ref.~\cite{S08}, two main mechanisms should be relevant for 
the dependence  of $\eta_I$ on $\lambda$: Shape fluctuations lead to energy 
dissipation that increases $\eta_I$, while alignment with the flow direction 
causes a decrease of $\eta_I$ with increasing $\lambda$. 
Our model, differently from the theory\cite{DM1,DM2} and other 
simulations,\cite{MIS10,Rah10,Zha13,Thi13,Kao14} includes thermal membrane 
undulations as well as thermal noise in the fluid.
The former effect is known to be relevant for the internal 
dynamics\cite{L12,A12} and the interaction between vesicles,\cite{Helf78} 
while the latter contribution induces Brownian diffusion of vesicles 
across the channel. This implies that a vesicle suspension with Brownian
motion can never attain the state of a regular array of vesicles arranged
periodically along the centerline of the channel, which is found as a state
of minimum dissipation in simulations without thermal noise.\cite{Thi13} 
A rough estimate of the importance of thermal effects can
be given considering the rotational Peclet number
$Pe=\eta_{out} \dot \gamma R_0^2/(k_B T)$, which is $Pe=12.0$ in our study.
In the case of a three-dimensional model of vesicles,  
it was found that thermal fluctuations cannot be neglected for rotational
Peclet numbers as large as $Pe=1200$.\cite{nogu05}
More work is in progress 
to elucidate the role of noise in the observed behavior of the intrinsic
viscosity.  Moreover, the formation of depletion layers next to the walls is
observed for concentrated suspensions: Their width increases with the viscosity 
contrast in the TT regime and then diminishes when entering the TU phase. 

In two dimensions, vesicles, capsules and red blood cells cannot be distinguished,
because shear elasticity has no analog in two dimensions. Indeed, simulations
in two dimensions have been employed to study the behavior of suspensions of
red blood cells in microcapillary flows.\cite{F07,F12} However, as far as the 
detailed dynamics of individual soft particles and its effects on the suspension
viscosity is concerned, the two-dimensional model more resembles fluid vesicles --- 
exactly because of the absence of shear elasticity in two dimensions. In order 
to elucidate the different behavior of vesicle and cell suspensions, simulations
in three dimensions are therefore essential.\cite{F11,G14,F14}

\end{document}